\documentclass[aps,longbibliography,preprint]{revtex4-1}

\usepackage{amsmath,bm}
\usepackage{graphicx,color}
\usepackage[update,prepend]{epstopdf}
\usepackage{enumitem}

\begin{document}
\title{Unexpected secondary flows in reverse nonequilibrium shear flow simulations}

\author{Antonia Statt}
\email{astatt@princeton.edu}
\affiliation{Department of Chemical and Biological Engineering, Princeton University, Princeton, New Jersey 08544, USA}

\author{Michael P. Howard}
\email{mphoward@utexas.edu}
\altaffiliation[Present address: ]{McKetta Department of Chemical Engineering, University of Texas at Austin, Austin,
Texas 78712, USA}
\affiliation{Department of Chemical and Biological Engineering, Princeton University, Princeton, New Jersey 08544, USA}

\author{Athanassios Z. Panagiotopoulos}
\affiliation{Department of Chemical and Biological Engineering, Princeton University, Princeton, New Jersey 08544, USA}

\newcommand{\vv}[1]{\mathbf{#1}}

\begin{abstract}
We simulated two particle-based fluid models, namely multiparticle collision dynamics and dissipative particle
dynamics, under shear using reverse nonequilibrium simulations (RNES).
In cubic periodic simulation boxes, the expected shear flow profile for a Newtonian fluid developed, consistent
with the fluid viscosities. However, unexpected secondary flows along the shear gradient formed when the simulation box was
elongated in the flow direction. The standard shear flow profile was obtained when the simulation box was longer
in the shear-gradient dimension than the flow dimension, while the secondary flows were always present when the flow dimension
was at least 25\% larger than the shear-gradient dimension.
The secondary flows satisfy the boundary conditions imposed by the RNES and have a lower rate of viscous dissipation in
the fluid than the corresponding unidirectional flows.
This work highlights a previously unappreciated limitation
of RNES for generating shear flow in simulation boxes that are elongated in the flow dimension, an important
consideration when applying RNES to complex fluids like polymer solutions.
\end{abstract}
\maketitle

\section{Introduction}
The reverse nonequilibrium simulation method (RNES) developed by M\"uller-Plathe
\cite{MuellerPlathe1997,Muller-Plathe1999}
is a well established technique for computing transport coefficients of particle-based fluid models. It was initially
proposed as a method to determine thermal conductivity \cite{MuellerPlathe1997} and was later extended to the
shear viscosity \cite{Muller-Plathe1999}. RNES has been used to calculate the thermal conductivity of simple
liquids \cite{MuellerPlathe1997,Reith2000},  salts \cite{Ding2017}, carbon  nanotubes \cite{Alaghemandi2009,Osman2001},
and silicon \cite{El-Genk2018}; to study the Ludwig--Soret effect \cite{Zhang2005,Reith2000}; and to measure the shear viscosities
of simple fluids \cite{Soddemann2003}, polymer solutions and melts \cite{Guo2002,Nikoubashman2017,Gonzalez2017},
ionic liquids \cite{Kelkar2007_2}, alcohols \cite{Kelkar2007}, and water \cite{Mao2012}.
RNES has also been successfully applied to more complex systems under shear, including colloidal suspensions of
nanoparticles \cite{Zhao2008,Sambasivam2018,Heine2010,Cerbelaud2017,Olarte-Plata2018}, surfactant solutions \cite{Meng2015},
and asphalt \cite{Yao2016}. More recently, RNES was used not only to compute transport properties, but also to
investigate dynamic phenomena like the shear-induced reorientation of diblock-copolymer lamellae \cite{Schneider2018} and the
aggregation of patchy particles in flow \cite{Mountain2017}.

The underlying idea of RNES is to impose an ``effect'' on a system in an unphysical way and measure
the ``cause''. In many nonequilibrium simulation techniques, a gradient (cause) is imposed and a flux (effect)
is measured \cite{Hoover:1983wb}, but RNES reverses this picture. For example, in RNES, stress can be
generated by an unphysical transfer of momentum between particles \cite{Muller-Plathe1999}, driving the system out of equilibrium
and resulting in a physical momentum flux. The stress gives rise to a corresponding flow profile that can
be measured. Given an imposed flux and measured gradient, the relevant transport coefficient
(e.g., shear viscosity) can be extracted within the linear-response regime.

RNES possesses many desirable properties of a nonequilibrium method \cite{Bordat2004,MuellerPlathe1997}.
It is quickly converging compared to equilibrium methods, such as the Green--Kubo relations \cite{Zwanzig1965}, to
determine transport
coefficients. RNES is compatible with a multitude of different particle-based models.
It also does not require a specific simulation box geometry and accommodates periodic
boundary conditions, which eliminates artificial wall effects that could cause measured fluid properties to differ
from the bulk in small, wall-bounded simulation boxes \cite{Perram:1975uq}. Additionally, RNES can be made to conserve momentum and energy and acts as its own
thermostat~\cite{Bordat2004}, avoiding the challenges of applying an external thermostat out of equilibrium
\cite{Evans:1986ug}.
Because of its straightforward implementation and flexibility, RNES is widely used and is implemented in many simulation
packages, including LAMMPS \cite{Plimpton1995}, HOOMD-blue \cite{Anderson2008}, OPENMD \cite{Louden2017}, and
ESPResSo \cite{Limbach2006}.

\begin{figure}
    \centering
    \includegraphics[width=12cm]{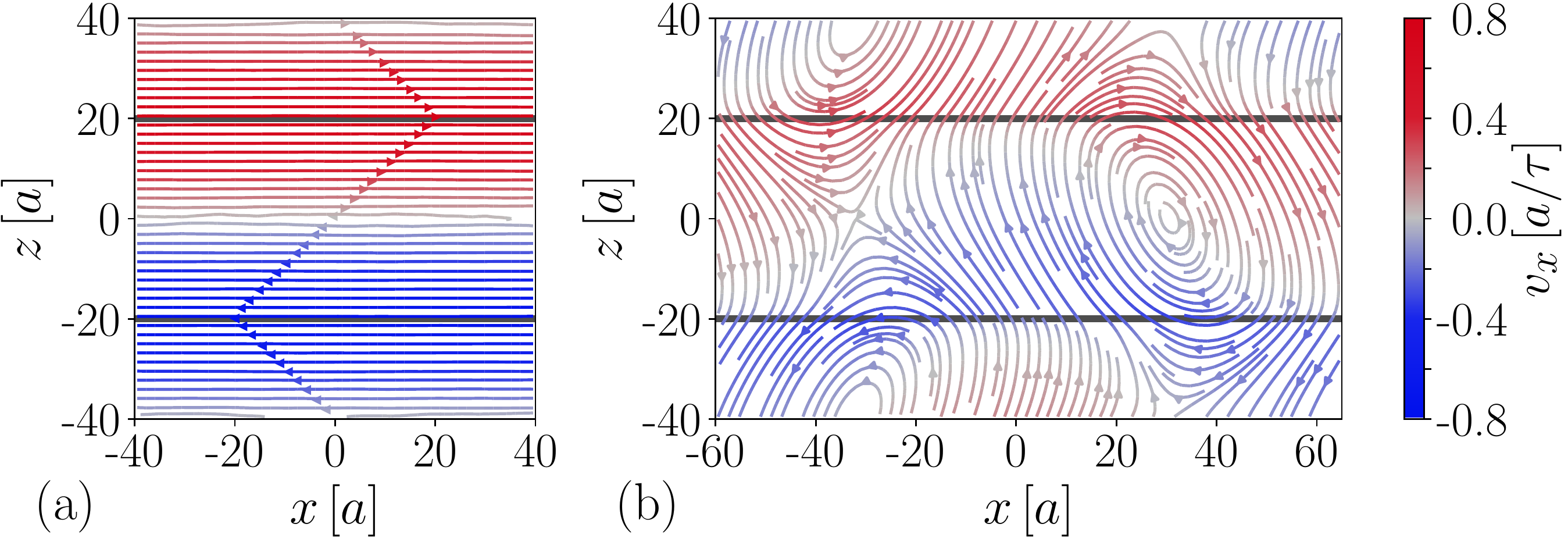}
    \caption{Streamlines in the $xz$-plane for the flow induced by RNES in (a) a cubic box ($80\,a\times80\,a\times80\,a$)
        and (b) an orthorhombic box ($125\,a\times 80\,a \times 80\,a$). Color indicates the $x$-component of the
        velocity, $v_x$. Gray horizontal lines mark the exchange slabs of width $w=1\,a$ at $z=\pm 20\,a$. }
    \label{fig:gamma_streamlines}
\end{figure}

Despite these many positive attributes, we have uncovered a previously unappreciated limitation of RNES
for simulating shear flow. We applied RNES to two different model fluids in various simulation box geometries.
Unexpectedly, we not only obtained the standard shear flows
(Fig.~\ref{fig:gamma_streamlines}a), but also more complicated flow patterns (Fig.~\ref{fig:gamma_streamlines}b)
in certain non-cubic boxes with periodic boundary conditions. In those cases, the flow had significant secondary components
along the shear gradient, and the shear stress did not have constant magnitude throughout the system. Such flows are
undesirable because they severely complicate the calculation of the shear viscosity using RNES.

In this article, we systematically interrogated the flow field, stress, and viscous dissipation as functions of
the simulation box geometry and the shear rate to determine \textit{why} and \textit{when} RNES did not generate
the expected shear flow. We show that the secondary flows emerge due to the periodic boundary conditions and are not
specific to the RNES method for generating shear flow, suggesting a hydrodynamic instability inherent to the flow and geometry.
The rest of the article is organized as follows. The RNES algorithm and its known
properties and features are explained in Sec.~\ref{sec:method}. The simulation details and fluid models
are described in Sec.~\ref{sec:simulations}. Our results are presented and discussed in Sec.~\ref{sec:results}
and are followed by our conclusions in Sec.~\ref{sec:conclusions}.

\section{RNES algorithm\label{sec:method}}
\begin{figure}
    \centering
    \includegraphics[width=9cm]{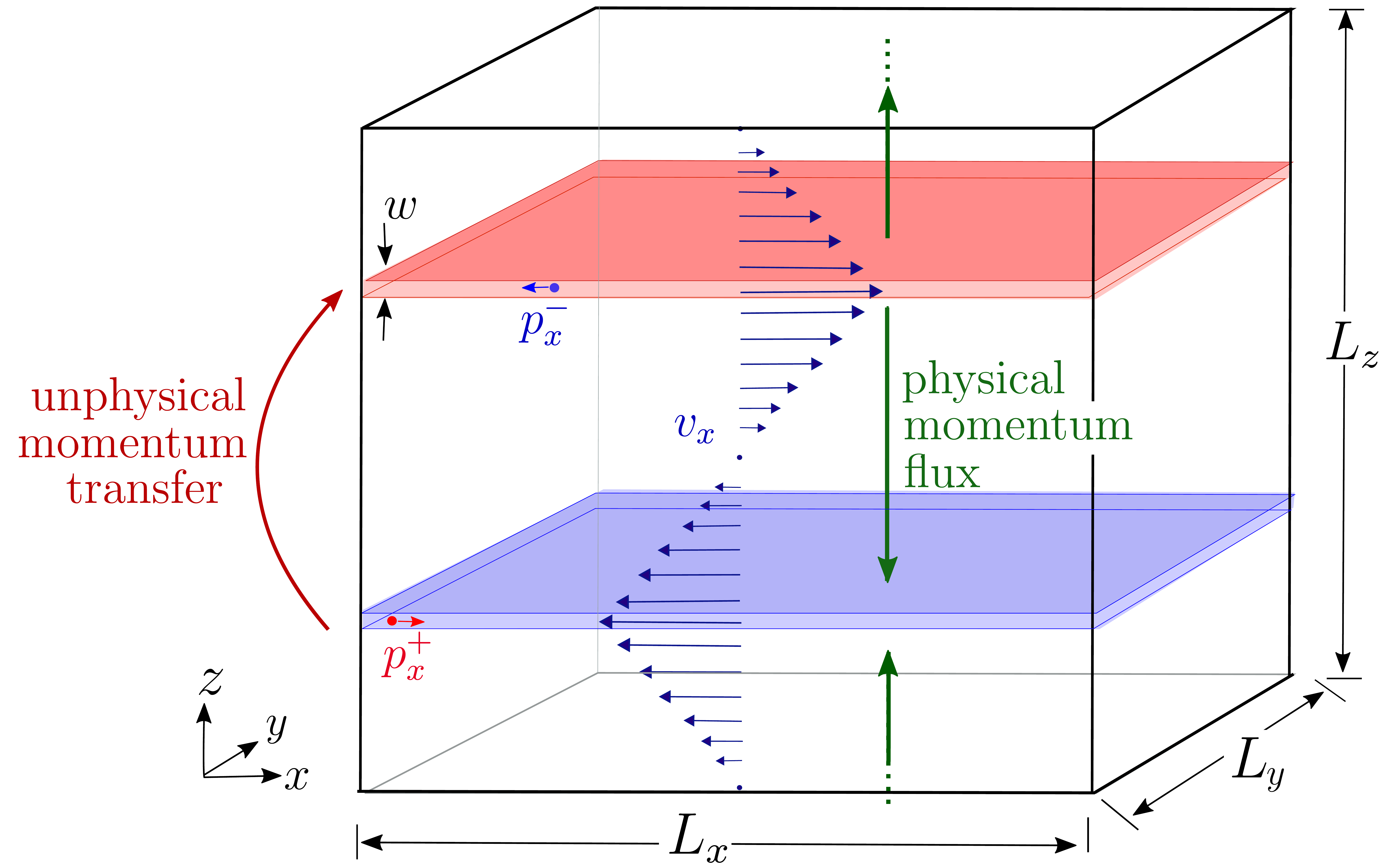}
    \caption{Sketch of the simulation box of dimensions $L_x \times L_y \times L_z$, where $x$ is the flow direction
        and $z$ is the gradient direction. The two exchange slabs of width $w$ are located at $-L_z/4$ (bottom, blue)
        and $+L_z/4$ (top, red). Unphysical momentum swaps between pairs occur in the top and bottom slab, driving a
        shear-induced physical momentum flux back (green). The expected velocity field given by eq.~\ref{eq:triangle},
        $v_x(z)$, is sketched in blue.}
    \label{fig:system_sketch_no_particles}
\end{figure}

We first summarize details of the RNES algorithm for simulating shear flow and its known features and limitations.
The simulation box is defined to be of size $L_x \times L_y \times L_z$, with $x$ being the
direction of flow and $z$ being the direction of the shear gradient (Fig.~\ref{fig:system_sketch_no_particles}).
The box is periodic in all three dimensions.
In the standard RNES algorithm for generating shear flow \cite{Muller-Plathe1999}, two exchange slabs of width $w$ are
constructed at $z = \pm
L_z/4$.
At regular intervals $\Delta t$ during the simulation, pairs of particles are selected from the slabs and their momenta
are swapped.
The particle with the most negative momentum in the flow direction, $p_x^-$, is selected from the upper slab,
while the particle with the most positive $x$-momentum, $p_x^+$, is chosen from the lower slab. An unphysical swap move
exchanges $p_x^-$ and $p_x^+$, resulting in a transfer of momentum $\Delta p_x = p_x^+ - p_x^-$. The exchange
of momentum generates a physical momentum flux with a corresponding shear stress, $\tau_{zx}$, at steady state
\cite{Muller-Plathe1999}:
\begin{align}
\tau_{zx} = \frac{\langle \Delta p_x \rangle}{2 L_x L_y \Delta t}.
\end{align}
Here, $\langle \Delta p_x \rangle$ denotes the average amount of momentum exchanged during the interval $\Delta t$,
and the factor of 2 is due to the periodic boundary conditions. If both particles in the swapped pair have the same mass,
the total momentum and kinetic energy of the system are conserved during the momentum transfer.

The imposed stress is expected to generate a flow field $\mathbf{v}$. Assuming the steady, incompressible flow of a
Newtonian fluid, the standard continuity equation and momentum balances governing the flow are \cite{Deen1998}
\begin{align}
\nabla \cdot \mathbf{v} &=0\label{eq:cont_eq}\\
\rho \mathbf{v} \cdot \nabla\mathbf{v}
    &= -\nabla p + \nabla \cdot \bm{\tau} \label{eq:momentum_balance} ,
\end{align}
where $\rho$ is the density, $p$ is the pressure, and $\bm{\tau} = \mu(\nabla\mathbf{v} + (\nabla\mathbf{v})^T )$ is the
viscous stress tensor with $\mu$ being the shear viscosity. We approximate the exchange procedure by a constant shear stress
$\tau_{zx}$ at the center of the exchange regions ($z = \pm L_z/4$) and a uniform pressure, which implies that there is no flow
in $y$ or $z$. Applying the periodic boundary conditions of the simulation box gives the steady-state shear flow field, $v_x(z)$,
\begin{equation}
v_x(z) = \begin{cases}
\dot{\gamma} (-L_z/2 - z),  & z < -L_z/4 \\
\dot{\gamma} z,  & |z| < L_z/4 \\
\dot{\gamma} (L_z/2 - z),  & z > L_z/4
\end{cases}\label{eq:triangle},
\end{equation}
where $\dot{\gamma} = \tau_{zx}/\mu$ is the shear rate in this geometry.
This flow field is illustrated in Fig.~\ref{fig:system_sketch_no_particles}.
Note that due to the periodic boundary conditions, the standard Couette flow ($|z| < L_z/4$)
is extended outside the exchange regions, resulting in an overall triangular profile.
In RNES, the width of the exchange regions $w$ should be as small as possible to minimally disturb
this expected flow field \cite{Schneider2018}, while still keeping the regions large enough to
contain a sufficient number of swapping pairs.

The imposed shear stress can be used to determine the shear viscosity $\mu$ from the
measured flow field within the linear-response regime,
\begin{equation}
\tau_{zx} = \mu \left\langle \frac{\partial v_x}{\partial z} \right\rangle
= \begin{cases}
\phantom{-}\mu \dot \gamma, & |z| < L_z/4 \\
-\mu \dot \gamma,   & |z| > L_z/4
\end{cases} .
\label{eq:stress}
\end{equation}
The shear rate $\dot{\gamma}$ can be varied by tuning the momentum exchange rate to change $\tau_{zx}$,
and $\mu$ is usually extracted from a series of measurements at different $\dot{\gamma}$.
RNES has technical limitations at both very low and very high shear rates.
For low shear rates, infrequent momentum swaps may not lead to a steady flow profile \cite{Tenney2010,Calderon2002}.
Analogous problems have been reported for the determination of thermal coefficients using RNES \cite{El-Genk2018}.
As suggested in Ref.~\citenum{Calderon2002},
this issue can be partially alleviated with a weaker, more frequent exchange of momentum,
e.g., by choosing pairs with momenta close to a target value instead
of the maximum \cite{Tenney2010}. This modification allows the flow rate to be tuned more precisely
\cite{Tenney2010,Petersen2010},
and also leads to a weaker system-size dependence and better convergence of the measured viscosity \cite{Petersen2010}.
For high shear rates, the velocities in the exchange regions deviate from the expected Boltzmann
distributions \cite{Tenney2010}. The velocity distribution in the lower slab develops a shoulder towards lower values,
whereas the distribution in the upper slab develops a shoulder towards higher values because the
imposed momentum transfer exceeds the system's ability to thermalize. In this case, RNES significantly
underpredicts viscosities relative to other computational methods \cite{Tenney2010}.
The numerical bounds on the accessible shear rates depend on the fluid model and must be determined
carefully by trial and error.

The RNES algorithm does not specify a box size or shape, and different geometries have been chosen by different
authors. For example, M\"{u}ller-Plate \cite{MuellerPlathe1997} originally used a box that was three times longer in
the gradient direction than in the flow direction ($L_z = 3 L_x$).
Nikoubashman and Howard chose $L_z = L_x$ to simulate polymer solutions in shear \cite{Nikoubashman2017}, while
Schneider et al. utilized boxes with $L_z = 2L_x-4.6L_x$ to study reorientation of diblock copolymers melts \cite{Schneider2018}.
Generally, and as is standard in molecular simulations \cite{Allen2017}, the simulation box must be large enough in all
dimensions to
avoid unphysical self-interactions through periodic boundaries, particularly between macromolecules like polymers.
Since many macromolecules align and stretch with the flow \cite{Doi1988}, the simulation box can similarly
be expanded along the flow dimension to match the expected deformation ($L_x > L_z$). This geometry
reduces the overall computational cost of the simulation compared to expanding the box in all dimensions.
However, as we will show in Sec.~\ref{sec:results}, elongating the box in the flow dimension has previously unknown
and undesirable consequences for RNES.

\section{Simulation Models\label{sec:simulations}}
We applied RNES to simulate shear flow for two fluids, one modeled using multiparticle collision dynamics (MPCD)
\cite{Malevanets1999,Gompper2009}
and the other using dissipative particle dynamics (DPD) \cite{Hoogerbrugge1992,Schlijper1995,Groot1997}. Both MPCD and DPD are
particle-based mesoscale models that faithfully resolve hydrodynamic interactions and incorporate the
effects of thermal fluctuations \cite{Noguchi2007}. In this article, we will describe the model parameters and our
results using $a$ as the unit of length, $\varepsilon$ as the unit of energy, and $m$ as the unit of mass. In this
system of units, $\tau = \sqrt{m a^2/\varepsilon}$ is the unit of time.

The MPCD fluid consisted of point-like particles of mass $m$. The particle positions and velocities were
propagated in alternating streaming and collision steps \cite{Malevanets1999,Gompper2009}. Particles moved
ballistically during the streaming step and were subsequently binned into cubic cells of size $a$. The particle
coordinates were collectively shifted by a random value between $\pm a/2$ during binning to ensure Galilean invariance
\cite{Ihle:2001ty,Ihle:2003bq}. In the ensuing collision step, the particle velocities relative to the cell-average
velocity were rotated by a fixed angle $\alpha$ around an axis randomly chosen from the unit sphere for each cell. This
collision procedure, also called stochastic rotation dynamics, conserves linear momentum in each cell. Because of the
local momentum conversation, it also conserves linear momentum globally.

The properties of the MPCD fluid are controlled by the particle number density, the rotation angle, the temperature,
and the time between collisions \cite{Malevanets1999,Gompper2009,Kapral2008}. We chose the density as $\rho=5\,a^{-3}$
and the rotation angle as
$\alpha=130^{\circ}$, and performed a collision every $0.1\,\tau$.
We additionally applied a Maxwell--Boltzmann rescaling thermostat \cite{Huang2015,Huang2010} to each cell to maintain a
constant temperature $T = 1.0\,\varepsilon/k_{\rm B}$ throughout the fluid, where $k_{\rm B}$ is Boltzmann's constant.
Since a thermostat is not required in RNES, we confirmed that removing the thermostat did not qualitatively change our
results.
The viscosity $\mu$ can be estimated for the MPCD fluid using kinetic theory \cite{Ripoll2005}, giving
$\mu=3.96\, \varepsilon \tau/a^3$ for our parameters.

The DPD fluid \cite{Hoogerbrugge1992,Koelman1993} was also represented by particles of mass $m$. A conservative force,
a dissipative force, and a random force acted between pairs of particles within a distance $r_{\rm c}$ of each other.
The dissipative and random forces were consistent with the fluctuation--dissipation theorem and
conserved local momentum. Details of the functional forms of the DPD interactions can be found in
Ref.~\citenum{Groot1997}.
We employed standard parameters: a maximum conservative force of $25\,\varepsilon/a$,
a drag coefficient of $4.5\,m/\tau$ for the dissipative force, a cutoff of $r_{\rm c}=1\,a$, and an
integration time step
of $0.01\,\tau$ \cite{Groot1997}. The particle number density was $\rho = 3\,a^{-3}$ \cite{Groot1997,Liu2015}, and the
temperature was $T = 1.0\,\varepsilon/k_{\rm B}$.

We determined the viscosity of the DPD fluid for our parameters using
RNES in cubic boxes with edge lengths ranging from $25\,a$ to $65\,a$, finding $\mu=0.87\,\varepsilon\tau/a^3$.
We independently validated the RNES measurement using the  periodic Poiseuille flow method~\cite{Backer2005}.
We divided the simulation box into two domains with $z < 0$ and $z\ge 0$ and applied a constant body force with
equal magnitude but opposite direction ($\pm x$) between the regions.  We extracted the viscosity by fitting the velocity profile to the expected parabolic
form~\cite{Backer2005}. The obtained value of $\mu=0.87\,\varepsilon\tau/a^3$  was in good agreement with the RNES
measurement and values reported elsewhere~\cite{Boromand2015}.

All simulations were performed on graphics processing units using HOOMD-blue (version 2.1.1)
\cite{Anderson2008,Glaser2015,Howard2018,Phillips2011}
with our own implementation of RNES. We additionally reproduced selected results for the MPCD fluid
using another
implementation \cite{ArashPrivateCommun} and for the DPD fluid using LAMMPS (11 Aug 2017) \cite{Plimpton1995,LAMMPS}.
An example LAMMPS script for simulating the DPD fluid is included as supplemental material \cite{SI}.

\section{Results and Discussion\label{sec:results}}
We first simulated the MPCD fluid in two boxes: a cubic box of size $80\,a\times 80\,a \times 80\,a$
and an orthorhombic box, elongated along the flow direction, of size $125\,a\times 80\,a \times 80\,a$.
The slab exchange width was $w = 1\,a$, and the swapped particles were chosen to have momenta closest to a
target value of $\pm 0.5\,ma/\tau$. We swapped $202$ pairs in the cubic box and $316$ pairs
in the orthorhombic box every step, resulting in the same imposed stress ($\tau_{zx} = 0.158\,\varepsilon/a^{3}$)
for both boxes.
The cubic box developed the expected triangular velocity profile (eq.~\ref{eq:triangle}) with extrema in the
exchange regions (Fig.~\ref{fig:gamma_streamlines}a). No flows occurred in either of the other
directions ($v_y = v_z = 0$). We extracted a viscosity of $\mu = 3.95\,\varepsilon\tau/a^3$ from
the measured flow field using eq.~\ref{eq:stress}, in quantitative agreement with the viscosity estimated from kinetic
theory.

We expected that applying the RNES method in the orthorhombic box should result in the same velocity profile
because the same stress was applied; surprisingly, a markedly different flow field
developed (Fig.~\ref{fig:gamma_streamlines}b). The average velocity in the exchange regions was still along $x$ as
expected, but substantial flows along the shear gradient ($z$) were also obtained, giving an
overall two-dimensional flow field ($v_y$ remained zero). A movie of the three-dimensional streamlines as they develop
from a quiescent fluid can be found in the supplemental material~\cite{SI}. The resulting streamlines exhibited two
vortices and two stagnation points, with one of each between the exchange slabs. The vortices and stagnation points
were
stationary during the accessible simulation time.
The DPD fluid under shear had the same qualitative behavior as the MPCD fluid,
indicating that this surprising flow field is not specific to the MPCD model.
We have extensively tested and verified that the momentum exchanged between the slabs is in agreement with the
expected value in all simulated systems and that all flow fields satisfy the
continuity equation (eq.~\ref{eq:cont_eq}). An example of this calculation can be found in Fig.~S1 \cite{SI}.

One possible explanation for the presence of vortices in the flow could be the emergence of turbulence in the Couette
flow. To estimate this effect, we defined a Reynolds number ${\rm Re} = \rho U L_z/4\mu$ based on the maximum velocity
in the exchange slabs $U$ and the half-width between the exchange slabs, $L_z/4$. It has been shown that
the transition from laminar to turbulent flow in planar Couette flow can occur for Reynolds numbers as low as ${\rm Re}
\approx 300$ \cite{Orszag1980,Bayly1988,Lundbladh1991}.
For Fig.~\ref{fig:gamma_streamlines}a, the Reynolds number is ${\rm Re} \approx 20$, which
is well-below the laminar--turbulent transition for the classical Couette flow. Expanding the box along $x$ does not
change Re, and so turbulence based on the flow is not expected for the orthorhombic box either. To exclude possible
finite-size effects, we performed a simulation in a box of size $125\,a \times 80\,a \times 125\,a$ at ${\rm Re} = 20$
as well as at  ${\rm Re} = 50$ and obtained only flows like Fig.~\ref{fig:gamma_streamlines}a.

As an additional stability test, we introduced two vortices and two stagnation points into a cubic box
($80\,a\times80\,a\times80\,a$)  and followed the evolution of the flow.
The vortices dissipated completely after $70000\,\tau$ and the flow returned to the expected
triangular flow profile. This shows that the expected shear flow is not metastable with respect to the vortices for the
cubic box. Likewise, we initialized the expected shear flow in an orthorhombic box
($125\,a \times 80\,a \times 80\,a$) and observed the emergence of vortices from the triangular flow on a similar
timescale. The final steady state was independent of how the fluid was initialized for both box shapes.

One of the benefits of using the RNES method is the ability to readily compute the shear viscosity from simulations
from a simple shear flow profile (eq.~\ref{eq:stress}). The secondary flows in the orthorhombic box effectively
prevent this calculation and are undesirable for simulating simple shear. In the next
sections, we investigate the local shear stress (Sec.~\ref{sec:shearstress}), the effect of boundary conditions
(Sec.~\ref{sec:bcs}), and the viscous dissipation in the fluid (Sec.~\ref{sec:visc}) to understand when and why
the secondary flows appear so that they can be avoided.

\subsection{Shear Stress~\label{sec:shearstress}}
\begin{figure}
    \centering
    \includegraphics[width=12cm]{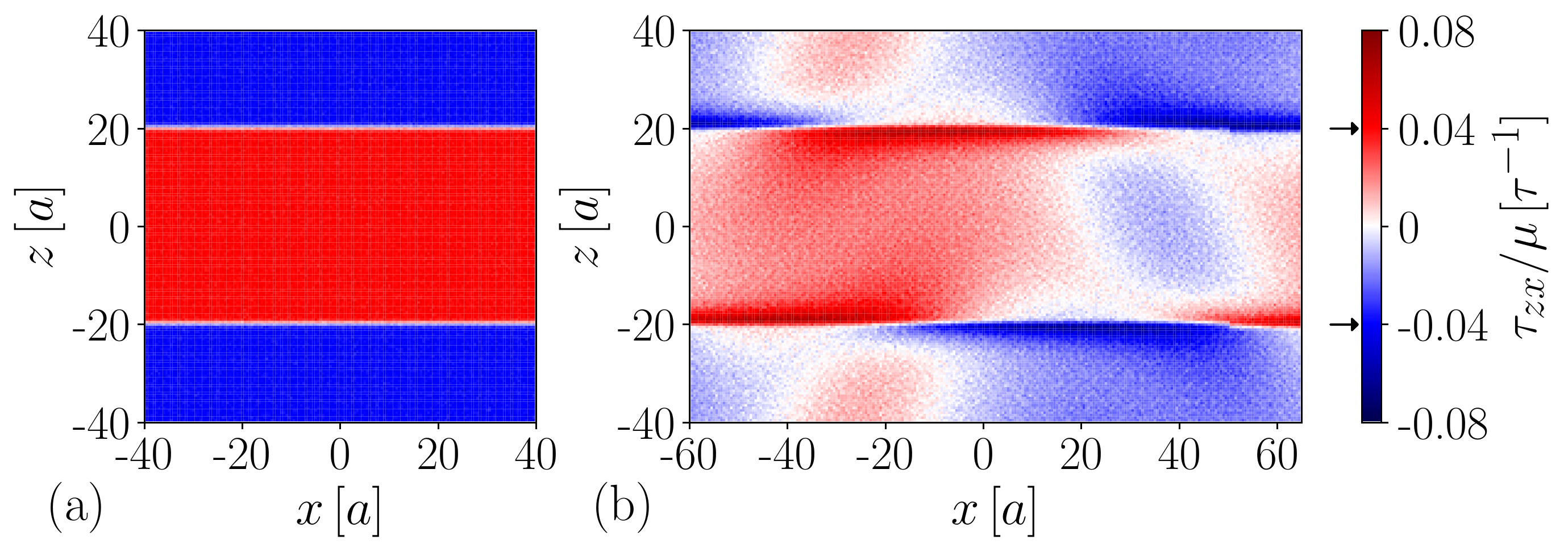}
    \caption{Local shear stress $\tau_{zx}/\mu$ in the fluid in (a) a cubic box and
        (b) an orthorhombic box for the flow fields of Fig.~\ref{fig:gamma_streamlines}. The expected
        values based on eq.~\ref{eq:stress}, $\tau_{zx}/\mu = \pm 0.04\,\tau^{-1}$, are indicated by
        arrows on the color bar.}
    \label{fig:shear_stress}
\end{figure}

In deriving the expected shear flow field (eq.~\ref{eq:triangle}), we assumed that the shear stress in the exchange regions was constant
and equal to the value imposed by RNES. In order to test this assumption, we computed the viscous stress tensor
$\bm{\tau}$ from the measured flow fields. Fig.~\ref{fig:shear_stress} shows the shear stress, $\tau_{zx}$, corresponding
to the flows in Fig.~\ref{fig:gamma_streamlines}. The cubic box (Fig.~\ref{fig:shear_stress}a) exhibited the expected step
function behavior (eq.~\ref{eq:stress}), $\tau_{zx}/\mu = \pm 0.04 \,\tau^{-1}$. Only the exchange regions of
width $w=1\,a$ at $z=\pm 20\,a$ deviated from the expected values, showing a shear stress close to zero. This is not surprising
because the stress in the fluid should vary continuously. In contrast, the shear stress in the orthorhombic box (Fig.~\ref{fig:shear_stress}b)
was strongly position dependent. The shear stress was lower near the vortices and stagnation points, but exhibited localized areas
of significantly higher stress close to the exchange regions.

In order to confirm that the width of the exchange region $w$ did not significantly affect the viscous stress, we systematically
decreased $w$ for both the MPCD and DPD fluids. Since the number of possible pairs to swap decreased with $w$,
we chose a lower target shear stress than in Fig.~\ref{fig:shear_stress}, using $\tau_{zx} = 0.068\,\varepsilon/a^3$
for MPCD and $\tau_{zx} = 0.015\,\varepsilon/a^3$ for DPD, which gives an expected shear rate of
$\dot{\gamma} = 0.0172\,\tau^{-1}$ for both fluids. The computed shear stress was averaged over $x$ to give
the average stress, $\langle \tau_{zx} \rangle$, as a function of $z$.

\begin{figure}
\centering
\includegraphics[width=12cm]{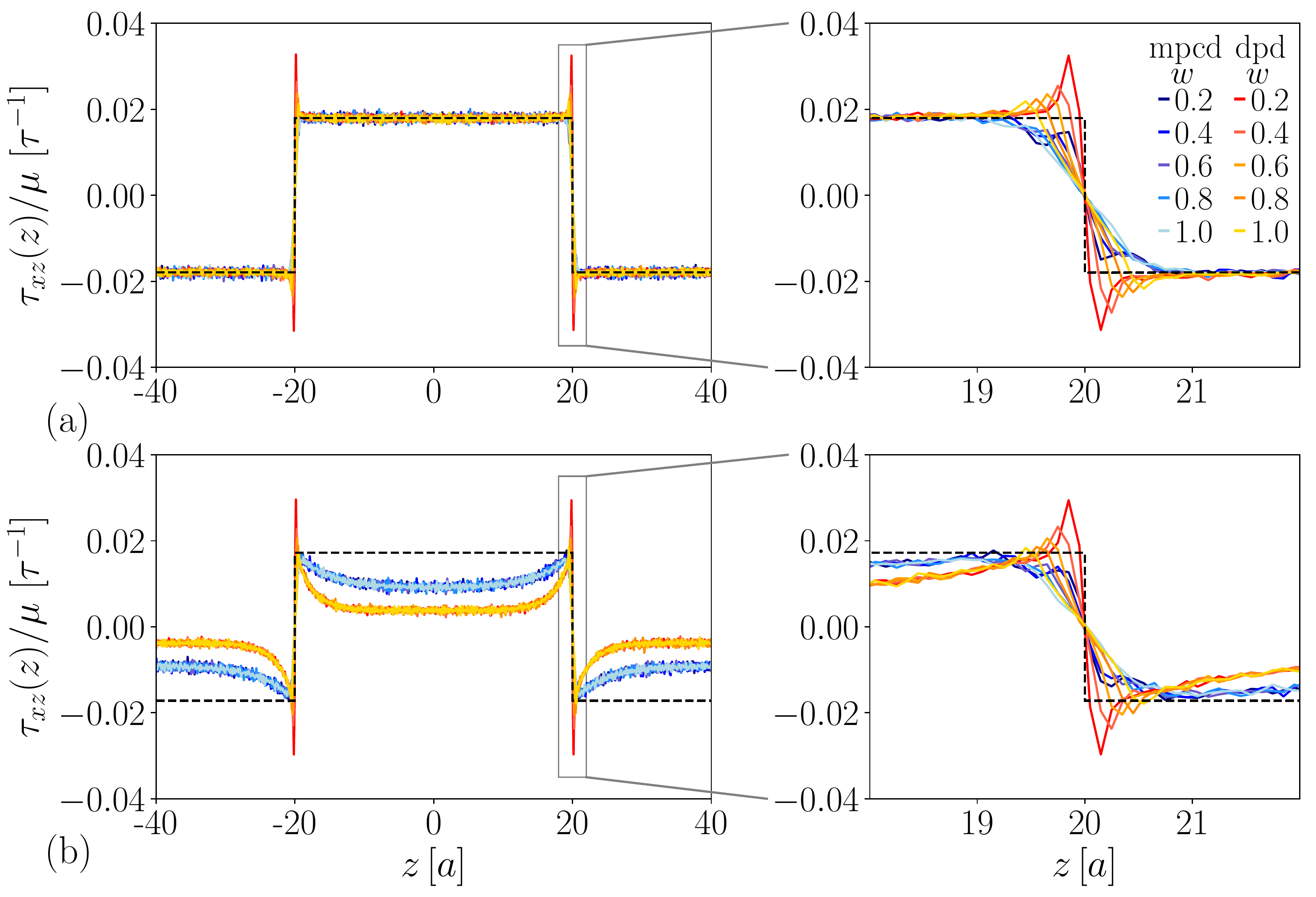}
\caption{Average shear stress $\langle \tau_{zx} \rangle$ in (a) a cubic box ($80\,a\times 80\,a \times 80\,a$) and
    (b) an orthorhombic box ($125\,a\times 80\,a \times 80\,a$) for the MPCD and DPD fluids. The width $w$ of the
    exchange region was varied from $0.2\,a$ to $1.0\,a$ as indicated in the legend. The expected step function,
    $\tau_{zx}/\mu = \pm 0.0172\,\tau^{-1}$, is shown as a dashed line.}
\label{fig:gamma_shear_stress_z}
\end{figure}

For the cubic box (Fig.~\ref{fig:gamma_shear_stress_z}a), $\langle \tau_{zx} \rangle$ was again a step function
for both fluids outside the exchange regions, as expected, regardless of the width of the exchange region, but
some deviations were observed near the exchange regions. The stress profile for the MPCD fluid was rounded
and increased in steepness with decreasing $w$, whereas the shear stress in the DPD fluid overshot the expected value
close to the exchange region by an amount that increased with decreasing $w$.
The overshoot in shear stress corresponds to the velocity being larger than expected near the exchange region,
which is consistent with previous simulations of DPD models \cite{Schneider2018}.
The rounded stress profile for the MPCD fluid is similarly consistent with rounding of the velocity profile
in the exchange region that has been reported elsewhere \cite{Cerbelaud2017}.

For the orthorhombic box (Fig.~\ref{fig:gamma_shear_stress_z}b), both fluids showed a significant reduction in
$\langle \tau_{zx} \rangle$ outside the exchange regions compared to the cubic box. As for the cubic box, the exchange
region width again did not influence the value of the stress in these regions. Also consistent with the cubic
box, the DPD fluid had stress overshoot in the exchange region, whereas the MPCD fluid had a rounded profile.
Most significantly, we note that $\langle \tau_{zx} \rangle$ in the exchange regions was essentially indistinguishable
between the cubic and orthorhombic boxes.
Although the shear stress in the exchange regions of the orthorhombic box was not \textit{constant}, the RNES
algorithm still imposed the same \textit{average} stress in the exchange regions for both simulation boxes.

\subsection{Boundary Conditions\label{sec:bcs}}
Our shear stress measurements suggest that the RNES method imposes boundary conditions that are inconsistent
with the assumptions used to derive eq.~\ref{eq:triangle}: namely, a constant shear stress $\tau_{zx}$ at $z = \pm L_z/4$
and a unidirectional flow ($v_y = v_z = 0$). In the previous section, we showed that RNES imposes the average
shear stress in the exchange regions. This is a weaker boundary condition than a constant stress
and does not enforce that $v_z = 0$. Clearly, $v_z \ne 0$ for the orthorhombic box (Fig.~\ref{fig:gamma_streamlines}b).

\begin{figure}
    \centering
    \includegraphics[width=12cm]{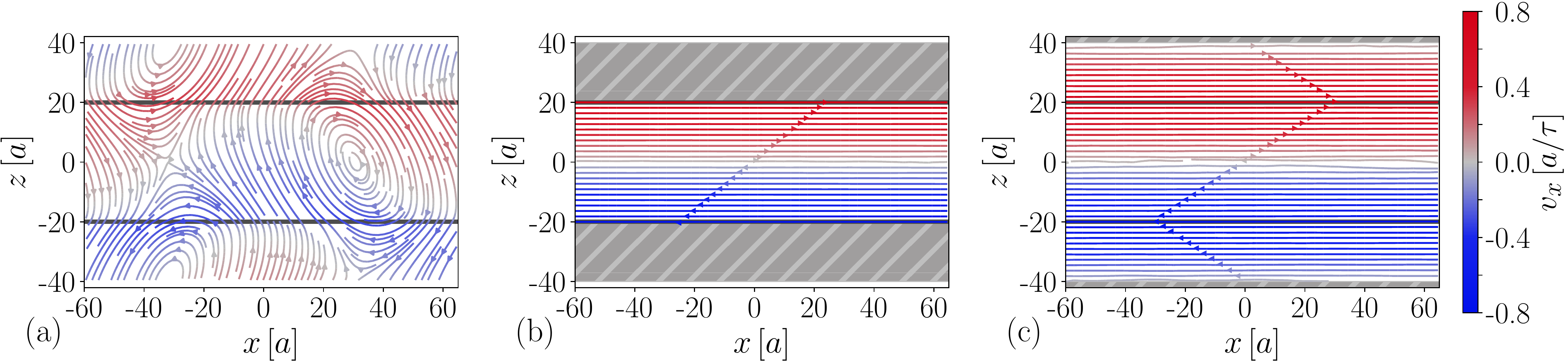}
    \caption{Streamlines in the $xz$-plane for the flow induced by RNES in
         (a) an orthorhombic box ($125\,a\times 80\,a \times 80\,a$) with periodic boundary conditions in all
         dimensions, (b) in the same box with slip walls placed at
         $z=\pm 20.5\,a$, and (c) in the same box with no-slip walls placed at $z=\pm 40\,a$. Color indicates the
         $x$-component of the
        velocity, $v_x$. Gray horizontal lines mark the exchange slabs of width $w=1\,a$ at $z=\pm 20\,a$. }
    \label{fig:gamma_streamlines_walls}
\end{figure}

To test the influence of including the condition that $v_z = 0$ in RNES, we inserted two smooth, hard walls into the
orthorhombic box just outside the exchange regions, effectively simulating half of the system ($|z| < L_z/4$).
We performed this test only for the MPCD fluid, which does not exhibit local structuring near a wall.
The MPCD particles were reflected from the hard wall using bounce-back reflections for a slip boundary condition.
The walls enforce that $v_z = 0$ since the fluid cannot penetrate the hard surface, but the slip boundary condition
does not modify the flow tangent to the surface. We then applied the RNES momentum
exchange to the fluid. We found that only the expected shear flow, $v_x(z)$, formed between the exchange regions, even
in boxes with $L_z > L_x$, as shown in Fig.~\ref{fig:gamma_streamlines_walls}b. From this test, we conclude that
secondary flows only form in RNES for fully periodic boxes
that are subject to an average stress boundary condition but do not constrain $v_z = 0$.

Because the previous test restricted the flow field to only one half of the RNES with periodic boundary conditions, we
also inserted two smooth, hard no-slip walls at exactly $z=\pm L_z/2$, effectively replacing the periodic boundary
condition in one dimension while keeping the rest of the system unchanged. We included virtual MPCD
particles~\cite{Bolintineanu2012,Lamura2001} at random positions in the walls with velocities drawn from the
Maxwell-Boltzmann
distribution to help enforce the no-slip condition. In this case, the entire
flow field given by eq.~\ref{eq:triangle} is recovered, as shown in Fig.~\ref{fig:gamma_streamlines_walls}c. To test
possible finite size effects, we
repeated the simulation in a box twice as big in $x$ and $z$ dimension and did not observe a difference in flow
behavior.

To test whether the secondary flows are due to the RNES momentum exchange procedure for generating the flow (average
stress condition), we considered an alternative nonequilibrium scheme.
We applied constant opposite body forces to all particles in the exchange regions instead of the usual RNES momentum
exchange method. Here, we applied a force $+\Delta f$ in the $x$-direction for each particle in the upper slab, and
$-\Delta f$ in the $x$-direction for each
particle in the lower slab. While the method of generating the flow was different, the resulting flow profiles and
stress distributions were identical to the ones observed by RNES. Together, these tests reveal that the
observed behavior is not unique to the nonequilibrium method by which the flow is created, but rather appears to be an
effect of the periodic boundary conditions.

\subsection{Viscous Dissipation\label{sec:visc}}
We have demonstrated that the unexpected secondary flows generated by RNES for simulation boxes that are elongated in
the flow dimension are consistent with the continuity equation and average momentum flux imposed by RNES and appear to
form due to the periodic boundary conditions in certain boxes. However, it is
still unclear {\it{when}} and {\it{why}} a certain flow field is realized for a given simulation box.
We use the total rate of viscous dissipation, $\Phi$, \cite{Deen1998}
\begin{equation}
\Phi[\mathbf{v}(\mathbf{r})] = \int {\rm d}\mathbf{r}~\bm{\tau}[\mathbf{v}(\mathbf{r})] : \nabla\mathbf{v}(\mathbf{r}),\label{eq:dissipation}
\end{equation}
as an order parameter to detect the transition from the expected shear flow to the regime
containing secondary flows. The rate of viscous dissipation is directly proportional to the rate of entropy production
in an isothermal, incompressible flow \cite{Deen1998,Iandoli2005}.
The expected viscous dissipation for the flow field given by eq.~\ref{eq:triangle}, $\Phi_0$, can be computed
analytically as $\Phi_0/V = \mu \dot{\gamma}^2$, where $V = L_x L_y L_z$ is the volume of the simulation cell.

To test if the rate of viscous dissipation detects the different flow profiles, we simulated the MPCD and DPD fluids in
rectangular boxes with aspect ratios varying from
$L_x/L_z=0.5$ to $2.7$.
We performed the simulations at both a fixed shear stress and a fixed maximum velocity according to
eq.~\ref{eq:triangle}. For the fixed shear stress simulations for the MPCD fluid, we varied $L_y = L_z$ from $80\,a$ to $125\,a$,
and kept the area $A = L_x L_y$ of the exchange region constant at $A=100^2\,a^2$.
The DPD fluid simulations at fixed shear stress were performed in boxes with a exchange slab area of $A=50^2\,a^2$
and varied $L_z$ from $30\,a$ to $65\,a$.
For the fixed maximum velocity simulations we picked boxes with $L_x=L_y=100\,a$ and varied $L_z$ from $50\,a$ to $125\,a$ for
the MPCD fluid. Similarly, the DPD fluid simulations were performed in boxes with  $L_x=L_y=50\,a$ and $L_z$ from
$28\,a$ to $60\,a$.
Averaged velocity profiles $\langle v_x \rangle$ and $\langle v_z \rangle$ can be found in
Figs. S2 and S3~\cite{SI}.

We then computed the total viscous dissipation $\Phi$ from the measured flow fields using eq.~\ref{eq:dissipation}.
As demonstrated in Sec.~\ref{sec:shearstress}, the simulated flow field
deviated from the theoretically expected profile near the exchange slabs even when no secondary flows occurred.
To better facilitate comparison between the simulations and theory, we excluded the exchange slabs of size $w=1\,a$
from both the theoretical and numerical calculations of $\Phi$. We confirmed that neglecting these regions did not
qualitatively change our findings, although it did lead to a small, constant shift of at most 5\% in the reported
values of $\Phi$.

\begin{figure}
    \centering
    \includegraphics[width=9cm]{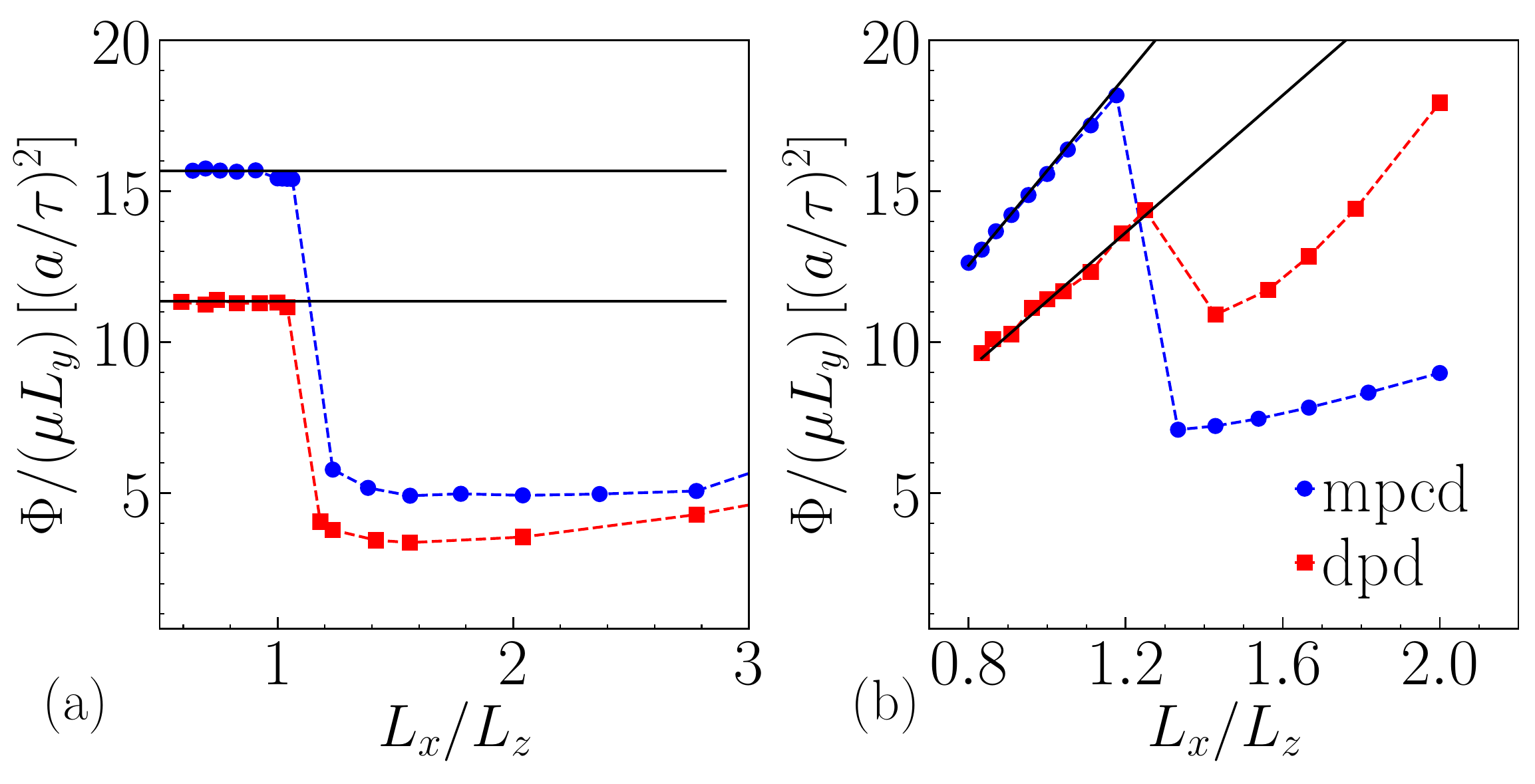}
    \caption{(a) Total viscous dissipation $\Phi$ as function of box shape for fixed average shear stress of $\tau_{zx}/\mu = 0.04\,\tau^{-1}$
        for the MPCD fluid and  $\tau_{zx}/\mu =0.0688\,\tau^{-1}$ for the DPD fluid. (b) Total viscous dissipation for fixed
        maximum velocity in the exchange regions. Note that dissipation in the RNES exchange regions has been excluded
        from $\Phi$ in both the numerical and theoretical calculations. }
    \label{fig:dissipation_box_ratio}
\end{figure}

We calculated the total viscous dissipation $\Phi$ for both the
MPCD and DPD fluids for
various different box aspect ratios $L_x/L_z$ at a constant average shear stress (Fig.~\ref{fig:dissipation_box_ratio}a)
and a constant maximum velocity, resulting in a varying shear stress (Fig.~\ref{fig:dissipation_box_ratio}b).
The viscous dissipation measured from the velocity gradients agreed
with the theoretical predictions for all cases where the box aspect ratio was smaller than approximately $L_x / L_z < 1.05$.
Above a critical value of approximately $L_x / L_z > 1.25$, the total viscous dissipation dropped significantly, concomitant
with the emergence of secondary flows. The flow field that is realized in the simulation appears to minimize the total
viscous
dissipation.

\begin{figure}
    \centering
    \includegraphics[width=9cm]{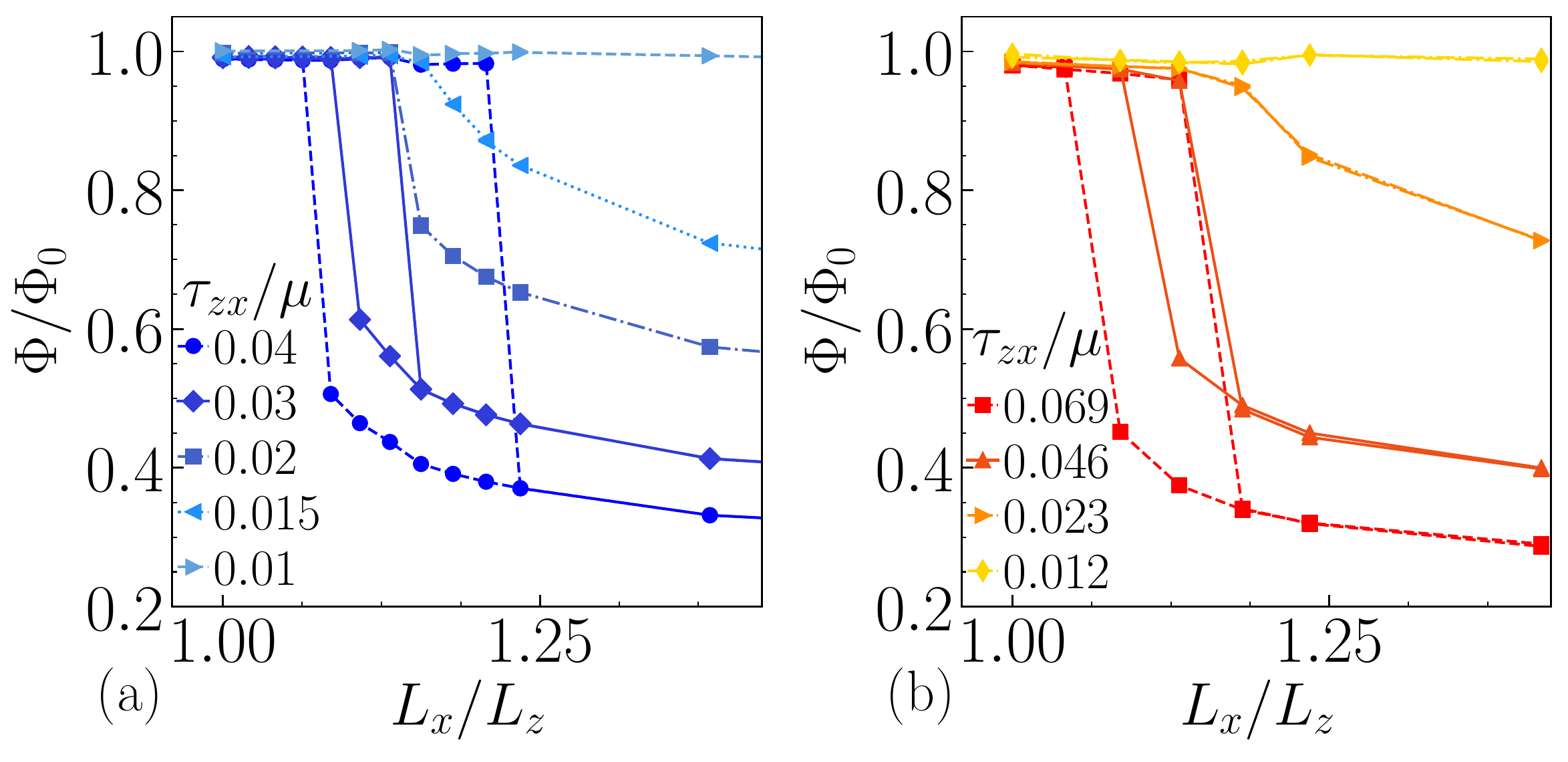}
    \caption{Ratio of measured to expected viscous dissipation, $\Phi/\Phi_0$, for the (a) MPCD fluid and
    (b) DPD fluid
        in simulation boxes of varied aspect ratio, $L_x/L_z$, and expected shear stress, $\tau_{zx}/\mu$. The shear rates are
        indicated in the legends.}
    \label{fig:dissipation_box_ratio_gamma}
\end{figure}

We subsequently investigated the shear rate dependence of the secondary flows using $\Phi$ as a parameter to identify
the two different flow field types. We simulated both fluids at
different momentum exchange rates in various box sizes and calculated $\Phi$ (Fig.~\ref{fig:dissipation_box_ratio_gamma}).
We found that the system showed significant hysteresis effects for box
aspect ratios $L_x/L_z$ between 1.05 and 1.25. We initialized two simulations for
each box geometry and shear rate: one with a flow field given by eq.~\ref{eq:triangle} and the other
with two vortices superimposed by an additional component $v_z(x)\sim \cos{(2 \pi x /L_x)}$.
For lower shear rates, the two different flows quickly converged to
the same viscous dissipation and the same flow profile. For the higher shear rates, the two systems did not converge
during the length of the
simulation run of $600\,000\,\tau$ for MPCD and $70\,000\,\tau$ for DPD. We therefore report both measured values of
the viscous dissipation in those cases.

For both fluids and all shear rates, secondary flows formed when $L_x / L_z > 1.25$.
The drop in the viscous dissipation was most pronounced for the biggest perturbations, even when scaled relative to $\Phi_0$.
The viscous dissipation varied only weakly for the lowest perturbation, but weak secondary flows still developed in the
elongated boxes, as shown in Fig.~S4 \cite{SI}. We were
unable to detect a lower critical shear
rate to avoid secondary flow within the simulation accuracy.

Because of the hysteresis, it was difficult to estimate a precise box aspect ratio for the emergence of the secondary flows.
As visible in Figs.~\ref{fig:dissipation_box_ratio} and \ref{fig:dissipation_box_ratio_gamma}, boxes which were cubic or elongated
in the gradient dimension ($L_x \le L_z$) always developed the expected shear flow. In contrast, simulations in boxes
with $L_x/L_z > 1.25$ always converged to a flow profile containing secondary flows.
Since it can be challenging to detect the secondary flows without careful sampling of the three-dimensional velocity field,
we advocate applying the RNES method for generating shear flow only when $L_x \le L_z$.

\section{Conclusions\label{sec:conclusions}}
In this work, we showed that the RNES method for simulating shear flow can generate unexpected secondary flows,
which do not allow a reliable calculation of the shear viscosity, in simulation boxes that are elongated in the flow dimension.
We demonstrated these flows occurred over a range of box sizes and shear stresses for two different fluid models,
revealing that the effect is general.
Although it was challenging to precisely identify the box geometries leading to secondary flows due to hysteresis effects,
we showed that no secondary flows formed during the accessible simulation time for aspect ratios, $L_x/L_z$, below 1.05,
while they always occurred for ratios above 1.25. Weak secondary flows were obtained even for very low shear rates,
indicating that there was no detectable lower threshold within the accuracy of the simulations.

The flow fields in both the cubic and elongated boxes were shown to be consistent with the same boundary condition on
the average stress imposed by RNES.
However, this boundary condition did not constrain the velocity along the shear gradient, $v_z$. When the RNES boundary
conditions were augmented by hard walls to obligate $v_z = 0$, the secondary flows were suppressed and the
expected Couette flow profile was recovered. This test demonstrated that the periodic boundary conditions of the simulation
box were needed to obtain the secondary flows.
For elongated simulation boxes, the fact that the RNES algorithm is compatible with periodic boundary conditions turns out
to not be a feature, but rather the origin of the undesired behavior.

We measured the velocity fields to calculate the
stresses and viscous dissipation. We showed that the viscous dissipation is a good predictor for the emergence
of secondary flows, with the total viscous dissipation dropping significantly compared to the theoretically expected
value when the secondary flows occurred.
The amount of reduction in viscous dissipation depended on the shear rate, with a greater
reduction relative to the expected value at higher shear rates. We hypothesize that the secondary flows occurred
because they minimized the viscous dissipation due to the flow, which is proportional to the rate of entropy
generation. We speculate this behavior may be due to an underlying hydrodynamic instability with respect to the
boundary conditions in this particular  geometry and  flow. The stability analysis of the Navier--Stokes equations
for RNES is left as an intriguing subject of future work.

This study reveals a previously unappreciated limitation of the RNES method. We found that
only boxes which are approximately cubic or elongated in the shear-gradient dimension can be reliably used to
simulate standard shear flow, especially for higher shear rates.
 This has important implications for
recent applications of RNES to study flow behavior of complex fluids like colloids \cite{Cerbelaud2017,Olarte-Plata2018,Mountain2017}
nanoparticles \cite{Zhao2008,Sambasivam2018,Heine2010}, or polymers \cite{Nikoubashman2017,Meng2015,Schneider2018}.
Both of the fluids tested here are frequently used as background solvents in these simulations. Although it may be tempting
to extend the simulation box in the flow dimension for computational efficiency, we advocate using only boxes with
$L_x \le L_z$ to reliably generate the expected  shear flow.

\begin{acknowledgments}
We thank Florian M\"uller-Plathe and  Howard Stone for insightful discussions.
We gratefully acknowledge use of computational resources supported by the Princeton Institute for Computational Science
and Engineering (PICSciE) and the Office of Information Technology's High Performance Computing Center and Visualization
Laboratory at Princeton University. Financial support for this work was provided by the Princeton Center for Complex
Materials, a U.S. National Science Foundation Materials Research Science and Engineering Center (award DMR-1420541), and
 the King Abdullah University of Science and Technology (KAUST) Office of Sponsored Research (OSR) under Award No.
OSR-2016-CRG5-3073.

\end{acknowledgments}

\bibliography{reverse_perturbation}

\end{document}